\documentclass[aps,prd,twocolumn,superscriptaddress,floatfix,nofootinbib]{revtex4}

\usepackage{color}
\usepackage[dvips]{graphicx}

\newcommand{\lsim}{\mathrel{\mathop{\kern 0pt \rlap
      {\raise.2ex\hbox{$<$}}}\lower.9ex\hbox{\kern-.190em $ \sim$}}}
\newcommand{\gsim}{\mathrel{\mathop{\kern 0pt
      \rlap{\raise.2ex\hbox{$>$}}}\lower.9ex\hbox{\kern-.190em $\sim$}}}
\newcommand{\beq}{\begin{equation}}
\newcommand{\eeq}{\end{equation}}

\newcommand{\be}{\begin{equation}}
\newcommand{\ee}{\end{equation}}
\newcommand{\beqarr}{\begin{eqnarray}}
\newcommand{\eeqarr}{\end{eqnarray}}

\begin{document}

\title{Observations of annual modulation in direct detection \\ of relic particles
and light neutralinos}
\thanks{Preprint numbers: ROM2F/2011/07 and DFTT 11/2011}

\author{P. Belli}
\affiliation{Istituto Nazionale di Fisica Nucleare, Sezione di Roma ÒTor VergataÓ, I-00133 Rome, Italy}
\author{R. Bernabei}
\affiliation{Istituto Nazionale di Fisica Nucleare, Sezione di Roma ÒTor VergataÓ, I-00133 Rome, Italy}
\affiliation{Dipartimento di Fisica, Universit\`a di Roma ÒTor VergataÓ, I-00133 Rome, Italy}
\author{A. Bottino}
\affiliation{Dipartimento di Fisica Teorica, Universit\`a di Torino, I-10125 Torino, Italy}
\affiliation{Istituto Nazionale di Fisica Nucleare, Sezione di Torino I-10125 Torino, Italy}
\author{F. Cappella}
\affiliation{Istituto Nazionale di Fisica Nucleare, Sezione di Roma, I-00185 Rome, Italy}
\affiliation{Dipartimento di Fisica, Universit\`a di Roma ÒLa SapienzaÓ, I-00185, Rome, Italy}
\author{R. Cerulli}
\affiliation{Istituto Nazionale di Fisica Nucleare, Laboratori Nazionali del Gran Sasso, I-67010 Assergi (Aq), Italy}
\author{N. Fornengo}
\affiliation{Dipartimento di Fisica Teorica, Universit\`a di Torino, I-10125 Torino, Italy}
\affiliation{Istituto Nazionale di Fisica Nucleare, Sezione di Torino I-10125 Torino, Italy}
\author{S. Scopel}
\affiliation{Department of Physics, Sogang University, 
Seoul, Korea, 121-742}
%

%
\date{\today}

\begin{abstract}
The long--standing model--independent 
annual modulation effect measured by the DAMA Collaboration, 
which fulfills all the requirements of a dark matter annual modulation 
signature, and the new result by the CoGeNT experiment that shows a similar 
behavior are comparatively examined under the hypothesis of a
dark matter candidate particle interacting with the detectors' nuclei by a coherent elastic process.
The ensuing physical regions in the plane of the dark matter--particle mass
versus the dark matter--particle nucleon cross--section are derived
for various galactic halo models and by taking into account the
impact of various experimental uncertainties.
 It is shown that the
DAMA and the CoGeNT regions agree well between each other and are well fitted by
a supersymmetric model with light neutralinos which satisfies all available
experimental constraints, including the most recent results from CMS and ATLAS at
the CERN Large Hadron Collider.
\end{abstract}
\preprint{DFTT}
\maketitle

\section{Introduction}
\label{intro}

An annual modulation effect, as expected from the relative motion of the Earth
with respect to the relic particles responsible for the Dark Matter (DM) in the
galactic halo  \cite{freese}, has been measured by the DAMA Collaboration since
long time  \cite{dama1997}, with an increasing exposure along 13 years
which, with the second generation DAMA/LIBRA apparatus, has reached
the value of 1.17 ton $\times$ year and a confidence level of 8.9 $\sigma$  \cite{dama2010}.

A very recent analysis of the data collected by the CoGeNT experiment
over a period of 442 days with a very low energy threshold Germanium
detector having a fiducial mass of 330 g has now led this
Collaboration to present the indication of a yearly signal modulation
at about 2.86 $\sigma$ \cite{cogent}.

The various experimental features required for a detector to be
sensitive to the expected annual modulation effect are not met by most
of the other direct detection experiments running at present.
However, it is intriguing that two of them (CDMS \cite{cdms} and
CRESST \cite{cresst}) found in their data some excesses of events over
what would be expected by them from backgrounds.  It is also
noticeable that, at least within one of the most widely considered
kind of DM particles, {\it i.e.} the one with an elastic coherent
interaction with the atomic nuclei of the detector material, the CDMS
and CRESST excess events would fall into (or close to) the physical
region singled out by the DAMA/LIBRA and CoGeNT annual--modulation
results.

The XENON100 Collaboration \cite{xenon100} and the CDMS Collaboration
(in re--analyses of their previous data \cite{cdmslow}) claim upper
bounds -- under a single set of fixed assumptions -- as in conflict
with the aforementioned results of the other experiments.  However,
problems related to the conclusions of Refs. \cite{xenon100,cdmslow},
as discussed in Refs. \cite{dama/xenon,collar/xenon} and in Ref.
\cite{collar/cdms}, respectively, and the existence of many
uncertainties in the procedures applied in the data handling by those
experiments, lead us to carry out here an analysis of the results of
Refs.  \cite{dama2010,cogent}, not conditioned by the results reported
in Refs.  \cite{xenon100,cdmslow}.

Though the model--independent annual modulation measured in the experiments of Refs.
\cite{dama2010,cogent} can be accounted for by a variety of
interaction mechanisms of relic particles with the detectors materials
\cite{dama2008}, we limit our analysis here to the case where the
signal is caused by nuclear recoils induced by elastic coherent
interactions with the DM particles. For simplicity as in a commonly
used nomenclature, in the following we will call a generic particle
with these features a WIMP (Weakly Interacting Massive Particle),
although the term WIMP identifies a class of DM particles which can
have well different phenomenologies, like {\em e.g.}  a preferred
interaction with electrons \cite{wimpele}.

Thus, in Sect. \ref{sec:region}, by using the results of Refs.  \cite{dama2010,cogent}
we first determine what are the physical regions pertaining to the DAMA and the
CoGeNT annual modulation data in terms of the WIMP mass and of the WIMP--nucleon
elastic cross--section at given confidence levels. In deriving these regions we take
into account the main origins of various experimental uncertainties, as well as
different forms for the Distribution Function (DF) of DM relic particles in the
galactic halo  \cite{bcfs}.

Subsequently in Sect.\ref{sec:lnm}
we show how the annual-modulation regions are well fitted by light neutralinos
within the effective Minimal Supersymmetric extension of the
Standard Model (MSSM) at the electroweak (EW) scale introduced in Ref.  \cite{lowneu}.
The relevance of light neutralinos in connection with the DAMA annual-modulation effect
was first discussed in Ref.  \cite{lowneudama}; their phenomenology  was then  developed in the context of
 direct  \cite{sum,discussing} and indirect  \cite{indirect} searches of DM particles.
The features of this specific realization of MSSM, dubbed  Light Neutralino Model (LNM), are also
 confronted here with the most recent constraints on supersymmetry (SUSY)
 derived at the Tevatron and at the Large Hadron Collider (LHC).

Particle physics models different from the LNM and potentially capable of generating light--WIMP particle candidates compatible with direct detection include
supersymmetric models which extend the MSSM by enlarging the particle field content, like in the next--to--minimal models \cite{NMSSM}, sneutrino dark matter models \cite{snu},
mirror--dark matter models \cite{mirror},
models with asymmetric dark matter \cite{asymm},
isospin--violating models \cite{isospin}, 
singlet dark matter models \cite{singlet},
specific realizations of grand unification \cite{GUT}, 
higgs--portal models \cite{portal}, composite models \cite{composite},
specific two--higgs doublet models \cite{specific};
secluded WIMPs \cite{secluded}.
Additional recent analyses can be found in Ref. \cite{others}.

Conclusions of our analysis are drawn in Sect. \ref{sec:con}.

\section{Regions related to the annual modulation effect in case of WIMPs}
\label{sec:region}

All experimental results discussed in the present Section are given in terms of plots in the plane
$m_{\chi}$ - $\xi\sigma_{\rm scalar}^{(\rm nucleon)}$, where
$\sigma_{\rm scalar}^{(\rm nucleon)}$ is
the WIMP--nucleon cross--section,
$\xi = \rho_{\chi} / \rho_0$; $\rho_0$ is the local total DM density
and  $\rho_{\chi}$ the local density of the DM candidate $\chi$. In the present section 
$\chi$ denotes a generic WIMP candidate, main responsible for the annual modulation effect
under discussion; it will specifically denote a neutralino in the Sections to follow.
The factor $\xi$ leaves open the possibility that the considered DM candidate does not provide the total
amount of local DM density.

\subsection{Phase--space distribution functions of dark matter}
\label{sec:df}

The quantity $\xi\sigma_{\rm scalar}^{(\rm nucleon)}$ can be derived from the experimental
spectra, once a specific DF is selected to describe the phase--space distribution function of
 the DM particle in the galactic halo. The appropriate form for the DF is still the subject of
extensive  astrophysical investigation. It is also possible that DM direct detection might be
 affected by  the presence of unvirialized components (see, for instance, Ref.  \cite{spergel}).
 Here we have taken a few samples of DFs, selected from the various realizations
 examined in Ref.  \cite{bcfs}, specifically:  i) the isothermal sphere (A0), ii) the Jaffe
 distribution (A4)  \cite{jaffe}, iii) a triaxial distribution (D2)  \cite{evans}
 (the notation adopted here follows those of Ref. \cite{bcfs}, to which we refer for
 further details). Notice that one could also have DF with a non-isotropic velocity dispersion
 (like distribution D2)
 and  co--rotating or counter--rotating halos. Then, though  the variety 
 of DFs discussed in this paper already offer a significant sample of DFs, this 
 selection is clearly not 
 (and could not be)  exhaustive of all possible situations. 
 
 As for the main parameters characterizing the various DFs (the local
 total DM density $\rho_0$ and the local rotational velocity $v_0$) we
 will take into account their physical ranges as discussed in Ref.
 \cite{bcfs}. Thus, we will take as representative values of $v_0$
 either one of the two extreme values or the central value of the
 physical range 170 km sec$^{-1} \leq v_0 \leq$ 270 km sec
 $^{-1}$. For each representative value of $v_0$ we take for
   $\rho_0$ either its minimal ${\rho_0}^{\rm min}$ or its maximal
   value ${\rho_0}^{\rm max}$, in the range compatible with the given value
   of $v_0$. As in Ref. \cite{bcfs}, ${\rho_0}^{\rm min}$
 (${\rho_0}^{\rm max}$) is defined as the value to be associated to
 $\rho_0$ when the visible mass provides its maximal (minimal)
 contribution to the total mass budget of the halo compatibly with
 observations. The numerical values for ${\rho_0}^{\rm min}$ and
 ${\rho_0}^{\rm max}$ depending on the DF and the values of $v_0$ will
 be taken from Table III of Ref. \cite{bcfs}. The escape
   velocity will be set at $v_{esc}$=650 km sec$^{-1}$.

\subsection{Annual-modulation regions in the considered model framework}
\label{sec:regions}

The about 9 $\sigma$ C.L. model independent positive results of the
DAMA/NaI and DAMA/LIBRA experiments \cite{RNC,ijmd,perflibra,dama2010}
and the recent positive hints by CoGeNT at 2.86 $\sigma$
C.L. \cite{cogent} can be analyzed in many corollary model--dependent
analyses. In all cases, many uncertainties in experimental parameters
as well as in necessary assumptions on various related astrophysical,
nuclear and particle-physics aspects must be taken into account.  In
the particular case of the WIMPs treated in this paper many sources of
uncertainties exist; some of them have been addressed {\it e.g.} in
Refs. \cite{RNC,ijmd,chan}. These affect all the results at various
extent both in terms of exclusion plots and in terms of allowed
regions/volumes and thus comparisons with a fixed set of assumptions
and parameters values are intrinsically strongly uncertain.  In the
following we will point out the effect of just one experimental
parameter, the quenching factor, whose precise determination is quite
difficult for all kinds of used detectors.

In fact, generally the direct measurements of quenching factors are
performed with reference detectors, and -- in some cases -- with
reference detectors with features quite different from the running
conditions; in some other cases, these quenching factors are not even
measured at all.  Moreover, the real nature of these measurements and
the used neutron beam/sources may not point out all the possible
contributions or instead may cause uncertainties because {\it e.g.} of
the presence of spurious effects due to interactions with dead
materials as {\it e.g.} housing or cryogenic assembling, if any;
therefore, they are intrinsically more uncertain than generally
derived. Thus, we specialize the present section to discuss the case
of the values of the quenching factor of Na and I in the highly
radiopure NaI(Tl) detectors of the DAMA experiments; analogous/similar
discussions should be pursued for the other cases.

As is widely known, the quenching factor is a specific property of the
employed detector(s) and not a general quantity universal for a given
material. For example, in liquid noble--gas detectors, it depends,
among other things, on the level of trace contaminants which can vary
in time and from one liquefaction process to another, on the cryogenic
microscopic conditions, etc.; in bolometers it depends for
  instance on specific properties, trace contaminants, cryogenic
conditions, etc. of each specific detector, while generally it is
assumed exactly equal to unity.  In scintillators, the quenching
factor depends for example on the dopant concentration, on
the growing method/procedures, on residual trace contaminants,
etc. and is expected to have energy dependence.  Thus, all these
aspects are already by themselves relevant sources of uncertainties
when interpreting whatever result in terms of DM candidates inducing
just recoils as those considered in the present paper.  Similar
arguments have already been addressed {\em e.g.} in
Refs. \cite{RNC,ijmd,chan,dama2010}. In the following, we will mention
some arguments for the case of NaI(Tl), drawing the attention to the
case of DAMA implications in the scenario considered in this paper.

The values of the Na and I quenching factors used by DAMA in the
corollary model--dependent calculations relative to candidates
inducing just recoils had, as a first reference, the values measured
in Ref. \cite{Psd96}. 
This measurement was performed with a small
NaI(Tl) crystal irradiated by a $^{252}$Cf source, by
applying the same method
previously 
employed in Ref. \cite{eijiri}. 
Quenching factors equal to (0.4 $\pm$
0.2) for Na and (0.05 $\pm$ 0.02) for I (integrated over the
$5-100$ keV and the $40-300$ keV recoil energy range,
respectively) were obtained.
 Using the same parametrization as in
Ref. \cite{eijiri}, DAMA measured in Ref. \cite{Psd96} quenching
factors equal to 0.3 for Na and 0.09 for I, integrated over the $6.5-97$
keV and the $22-330$ keV recoil energy ranges, respectively.  The
associated errors derived from the data were quoted as one unity in
the least significative digit.  Then, considering also both the large
variation available in the literature (see {\em e.g.}  Table X
of Ref. \cite{RNC}) and the use of a test detector \cite{RNC}, a
20\% associated error has been included. 
Nevertheless, some recent considerations, as those reported in Ref.
      \cite{tretyak} about the energy dependence of quenching factors for
      various recoiling ions in the same detector, have called our attention
      to the fact that the large uncertainties in the determination of
      Ref. \cite{eijiri} could be due, in a significant part, to 
      uncertainties in the parametrization itself, which we also adopted. 
      Another uncertainty could arise from the determination of integrated 
      values, while an increase of the quenching factor values towards lower
      energies could be expected, as observed in some crystal detectors as
      for instance CsI.

An additional argument on uncertainties on quenching factors in 
         crystals, and specifically for NaI(Tl), is the presence and the 
         amount of the well known channeling effect of low energy ions
         along the crystallographic axes and planes of NaI(Tl) crystals.
         Such an effect can have a significant impact in the corollary model 
         dependent analyses, in addition to those uncertainties discussed
         above and later on, since a fraction of the recoil events would    
         have a much larger quenching factor than that derived with neutron
         calibrations.
         Since the channeling effect cannot be generally put into evidence 
         with neutron measurements, as discussed in details in Ref. 
         \cite{chan}, only theoretical modeling has been produced up to now.   
        In particular, the modeling of the channeling effect described by  
        DAMA in Ref. \cite{chan} is able to reproduce the recoil spectrum
        measured at neutron beam by some other groups \cite{othergroup}.
        For completeness, we mention alternative channeling models, as that
        of Ref. \cite{Matyu08} where larger probabilities of the planar
        channeling are expected. Moreover, we mention the analytical
        calculation claiming that the channeling effect holds for recoils
        coming from outside a crystal and not from recoils produced inside   
       it, due to the blocking effect \cite{gelmini}.
       Nevertheless, although some amount of blocking effect could be   
       present, the precise description of the crystal lattice with dopant 
       and trace contaminants is quite difficult and analytical calculations 
       require some simplifications which can affect the result.

Recently, Ref. \cite{tretyak} pointed out the possibility that the quenching factors for nuclear
recoils in scintillators can be described with a semi--empirical formula having
only one free parameter: the Birks constant, $k_B$, which depends on the specific
set--up. Applying this procedure to the DAMA detectors operating underground and fixing the $k_B$
parameter to the value able to reproduce the light
response to alpha particles in these detectors, the expected Na and I quenching factors
are established as a function of the energy with values ranging from 0.65 to 0.55
and from 0.35 to 0.17 in the $2-100$ keV electron equivalent energy interval, for Na and I
nuclear recoils, respectively; as evident, also an energy dependence is pointed out there.

In the following analysis, we  present some of the many possible model--dependent analyses of 
the DAMA results, including at least some of the present uncertainties.
In particular, the uncertainties due to the description of the halo are accounted for some of 
the many possible halo models; we employ here the DFs mentioned in Sect.\ref{sec:df}.

\begin{figure*}[t]
\includegraphics[width=0.45\textwidth] {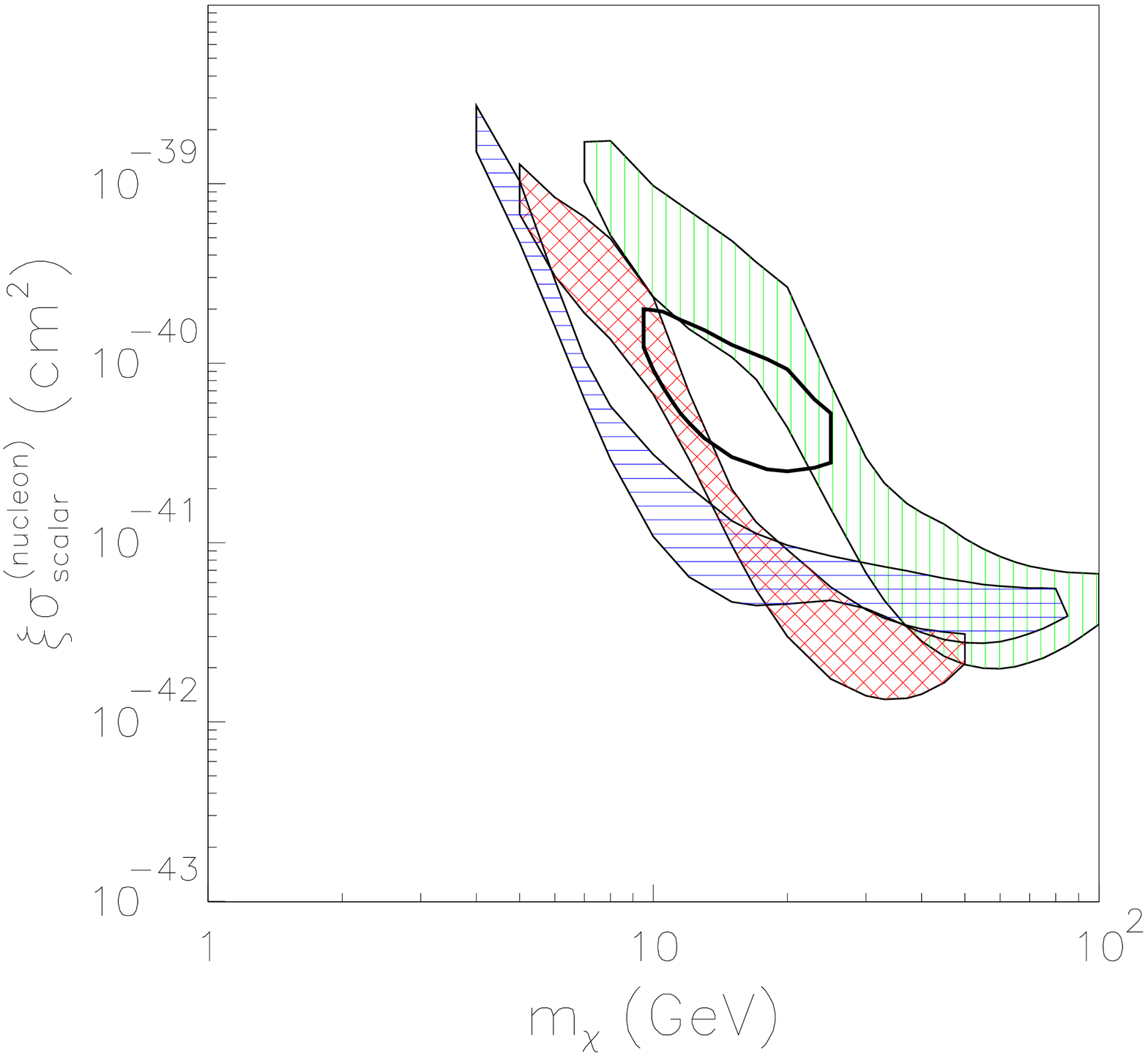}
\includegraphics[width=0.45\textwidth] {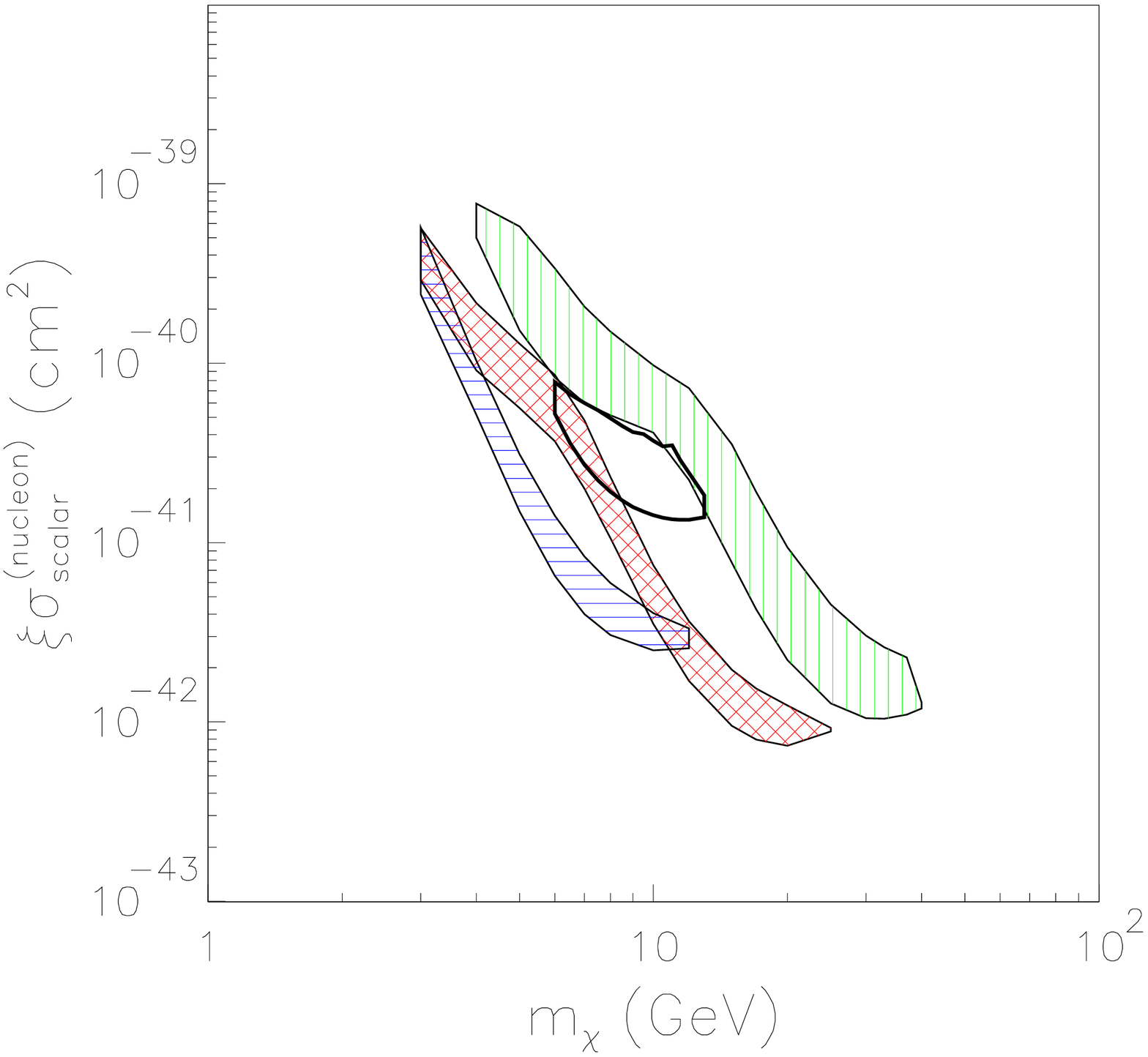}
  \caption{$\xi\sigma_{\rm scalar}^{(\rm nucleon)}$ as a function of
    the mass $m_{\chi}$ of a generic DM particle which interacts with
    nuclei by an elastic coherent scattering.  The halo DF is taken to
    be given by the isothermal sphere ((A0) in the notations of
    Subsect.\ref{sec:df} and Ref.\cite{bcfs}). The parameters are: i)
    in the left panel, $v_0 = 170$ km sec$^{-1}$, $\rho_0 = 0.18$ GeV
    cm$^{-3}$ ii) in the right panel, $v_0 = 270$ km sec$^{-1}$,
    $\rho_0 = 0.45$ GeV cm$^{-3}$ (see text for further details).  The
    three (colored) hatched regions denote the DAMA annual modulation
    regions, under the hypothesis that the effect is due to a WIMP
    with a coherent interaction with nuclei and in 3 different
    instances: i) without including the channeling effect ((green)
    vertically--hatched region), ii) by including the channeling effect
    according to Ref.\cite{chan} ((blue) horizontally--hatched region),
    and iii) without the channeling effect but using the
    energy--dependent Na and I quenching factors as established by the
    procedure given in Ref. \cite{tretyak} ((red) cross--hatched
    region).  They represent the domain where the likelihood--function
    values differ more than 7.5 $\sigma$ from the null hypothesis
    (absence of modulation).  The (non--hatched) region denoted by a
    (black) solid contour is the allowed region by the CoGeNT experiment
    when considering the modulation result given in Ref. \cite{cogent} 
    and the assumptions given in the text for
 the quenching factor and the form factor. This region is meant to include configurations whose
    likelihood--function values differ more than 1.64 $\sigma$ from the
    null hypothesis (absence of modulation). This corresponds roughly
    to 90\% CL far from zero signal.  In fact due to the presently
    more modest C.L. (about 2.9 $\sigma$) of this result with respect
    to the 9 $\sigma$ C.L. of the DAMA/NaI and DAMA/LIBRA evidence for
    Dark Matter particles in the galactic halo, no region is found if
    the stringent 7.5 $\sigma$ from absence of modulation is required
    as for DAMA.  It is worth noting that, depending on other possible
    uncertainties not included here, the channeled (blue) region could
    span the domain between the present channeled region and the
    unchanneled one.}
\label{fig1}
\end{figure*}

\begin{figure*}[t]
\includegraphics[width=0.45\textwidth] {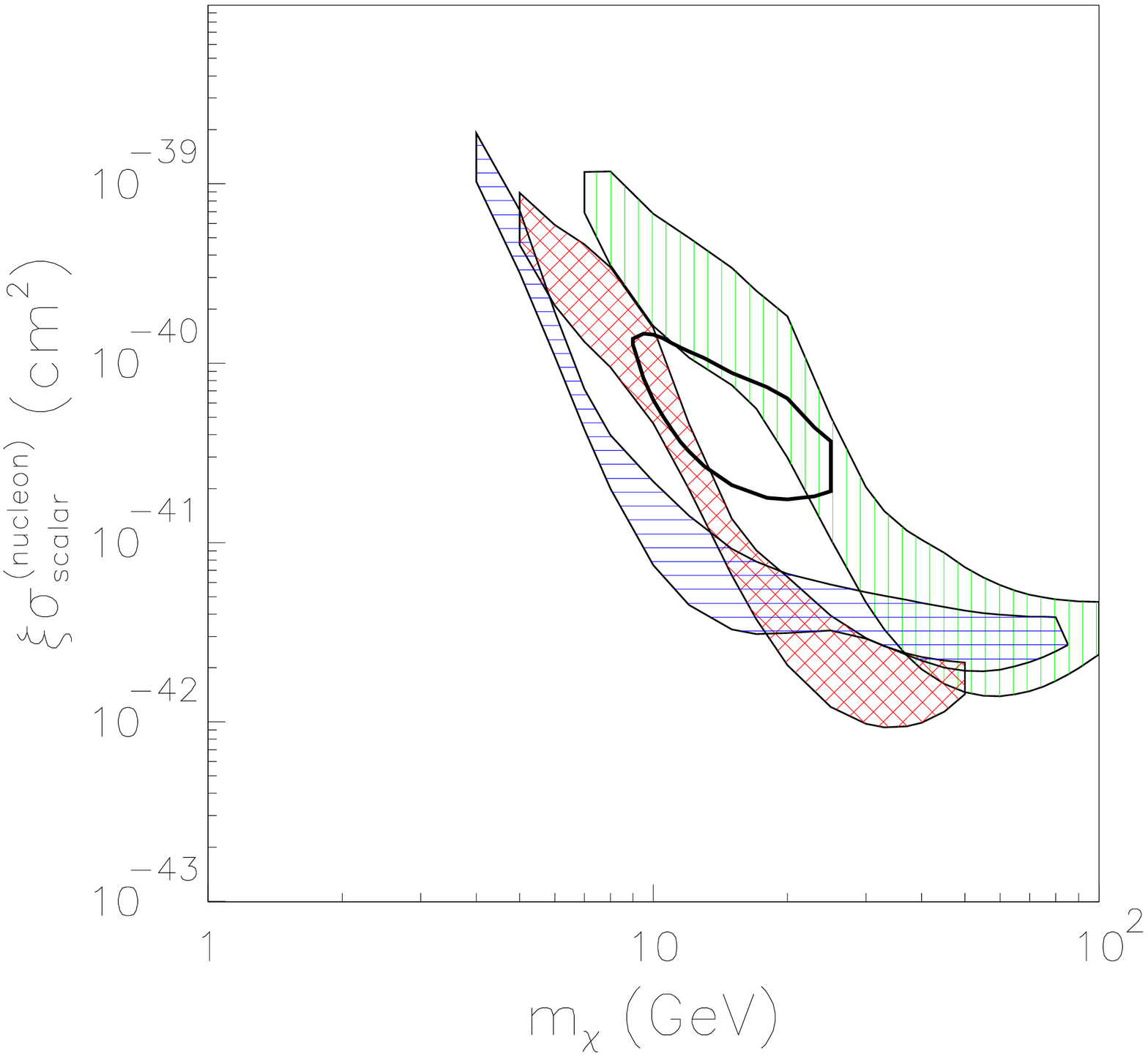}
\includegraphics[width=0.45\textwidth] {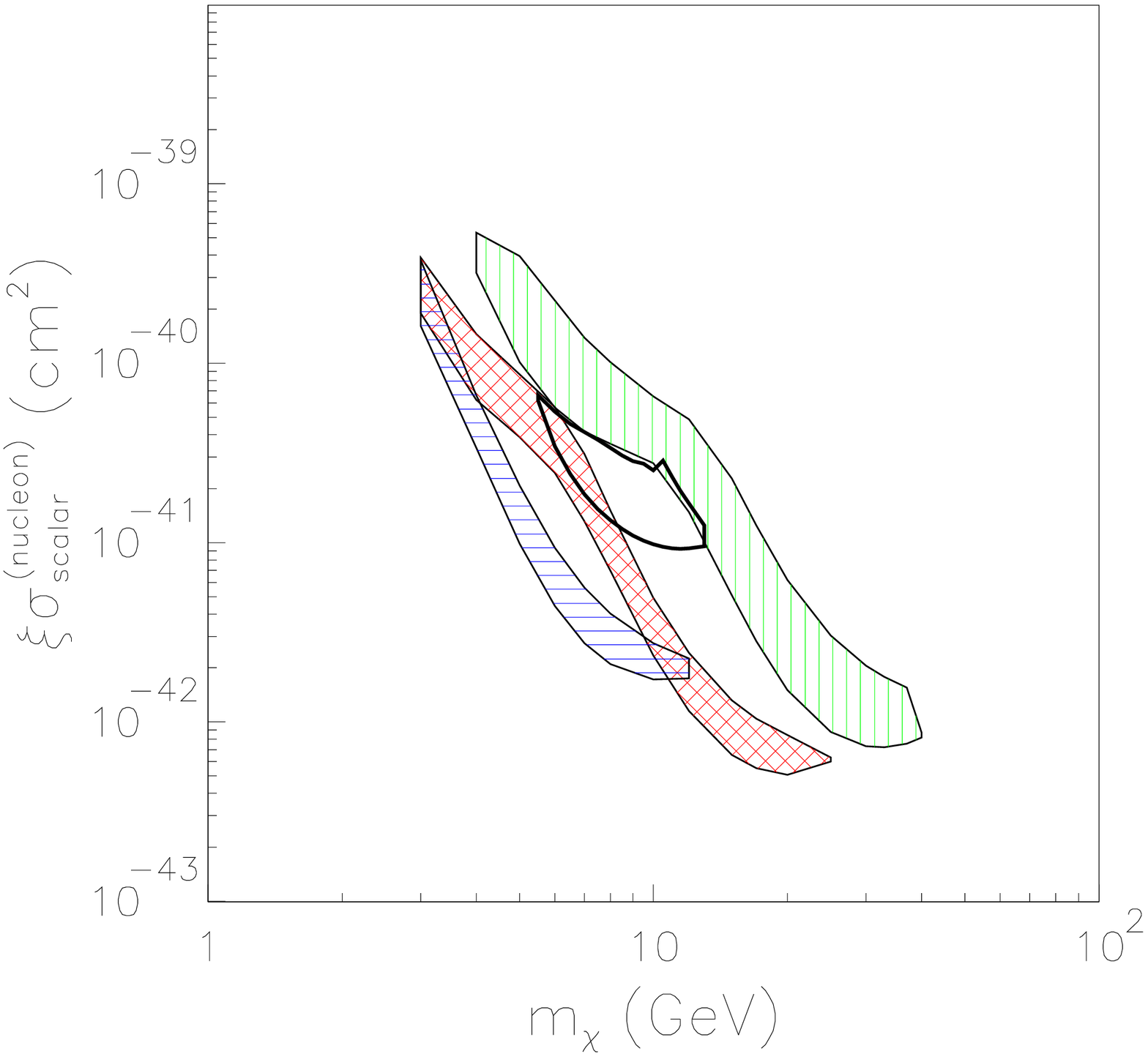}
\caption{As in Fig. \ref{fig1} except that here the halo DF is taken to be given by the Jaffe distribution 
\cite{jaffe} ((A4) in the notations of Subsect.\ref{sec:df} and Ref.\cite{bcfs}). The parameters are:
i)  in the left panel, $v_0 = 170$ km sec$^{-1}$, $\rho_0 = 0.26$ GeV cm$^{-3}$
ii) in the right panel, $v_0 = 270$ km sec$^{-1}$,  $\rho_0 = 0.66$ GeV cm$^{-3}$}  
\label{fig2}
\end{figure*}

\begin{figure*}[!ht]
\includegraphics[width=0.45\textwidth] {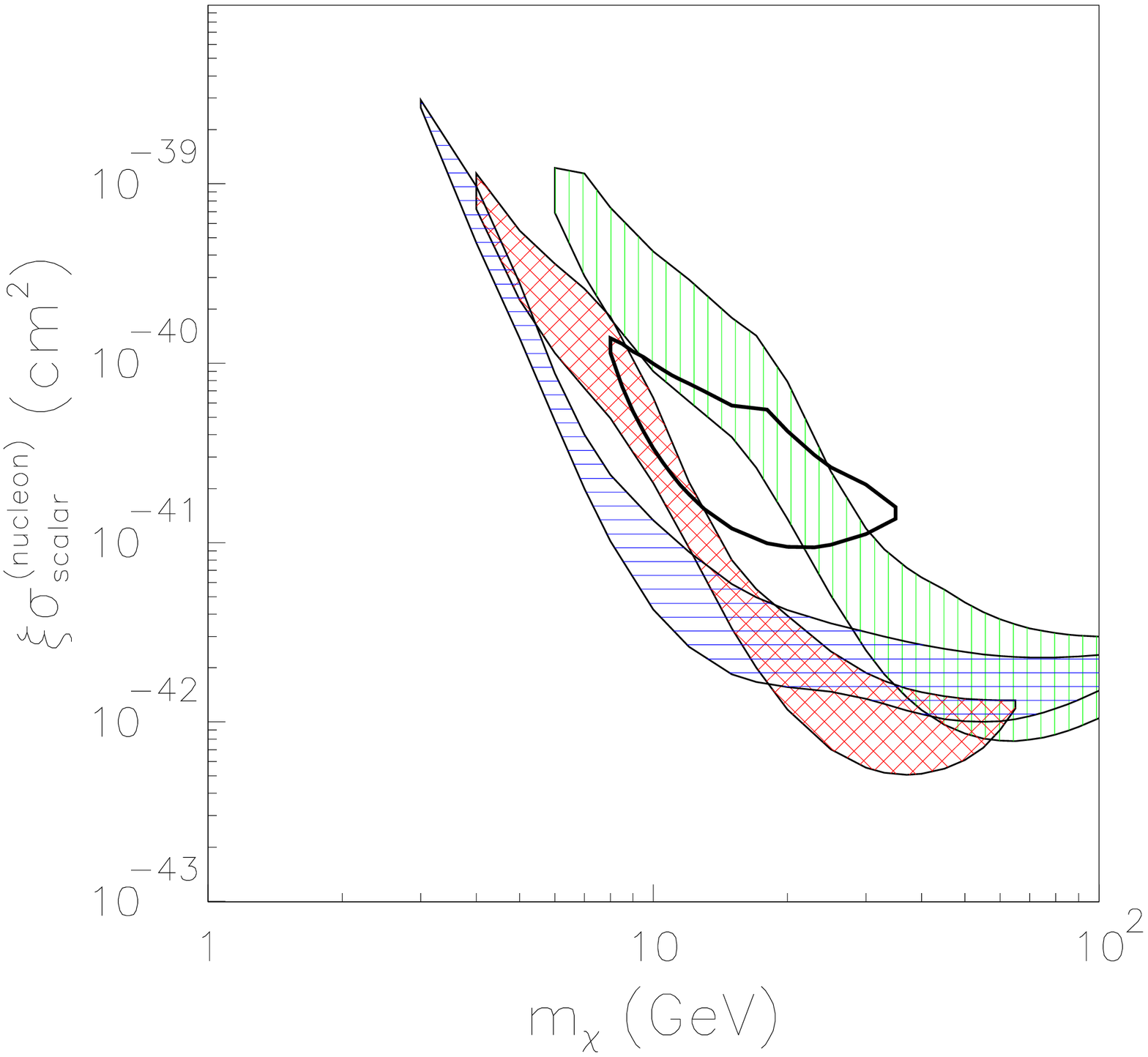}
\includegraphics[width=0.45\textwidth] {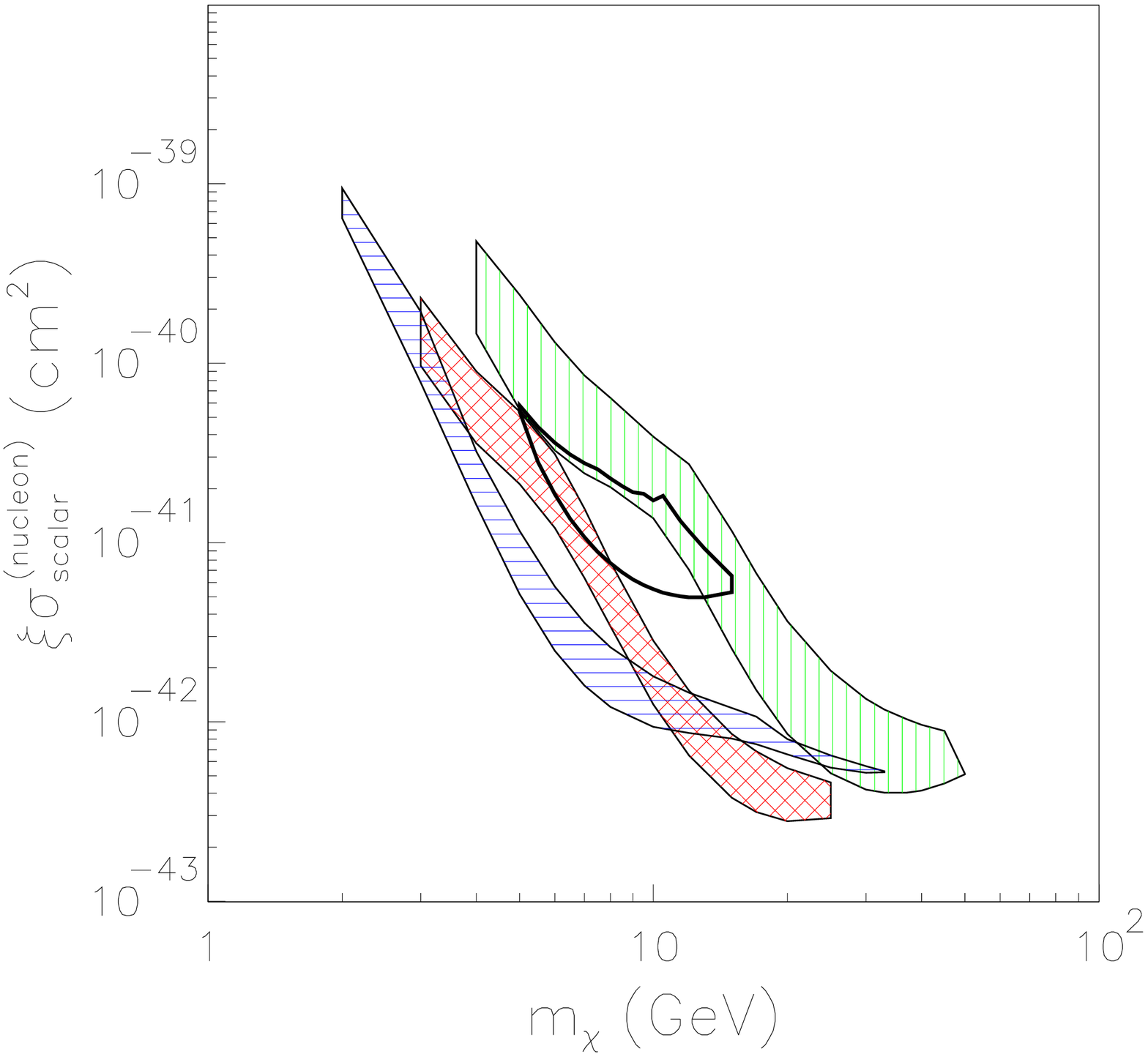}
\caption{As in Fig. \ref{fig1} except that here the halo DF is taken to be given by a triaxial distribution
\cite{evans} ((D2) in the notations of Subsect.\ref{sec:df} and Ref.\cite{bcfs}). The parameters are:
i)  in the left panel, $v_0 = 170$ km sec$^{-1}$, $\rho_0 = 0.50$ GeV cm$^{-3}$
ii) in the right panel, $v_0 = 270$ km sec$^{-1}$,  $\rho_0 = 1.27$ GeV cm$^{-3}$}
\label{fig3}
\end{figure*}

Figs. \ref{fig1}, \ref{fig2} and \ref{fig3} show $\xi\sigma_{\rm
  scalar}^{(\rm nucleon)}$ as a function of the Dark Matter particle
mass $m_{\chi}$ for the A0, A4 and D2 halo models \cite{bcfs}.  In
order to have a significative sample in terms of the physical ranges
of the relevant astrophysical parameters we have chosen to
display the annual--modulation regions for the two extreme values of
the local rotational velocity $v_0$, {\it i.e.}: $v_0 = 170$ km
sec$^{-1}$ (in the left panel of each figure) and $v_0 = 270$ km
sec$^{-1}$ (in the right panel).  In Fig. \ref{fig1}, where the case
of the A0 distribution function is shown, we have set the local total
DM density $\rho_0$ to be equal to its minimal value, $\rho_0 =
{\rho_0}^{\rm min}$, compatibly with the value of $v_0$, then $\rho_0
= 0.18$ GeV cm$^{-3}$ for $v_0 = 170$ km sec$^{-1}$ (in the left
panel) and $\rho_0 = 0.45$ GeV cm$^{-3}$ for $v_0 = 270$ km sec$^{-1}$
(in the right panel).  In Fig. \ref{fig2}, where we display the case
of the A4 distribution function, we set again $\rho_0 = {\rho_0}^{\rm
  min}$, then $\rho_0 = 0.26$ GeV cm$^{-3}$ for $v_0 = 170$ km
sec$^{-1}$ (in the left panel) and $\rho_0 = 0.66$ GeV cm$^{-3}$ for
$v_0 = 270$ km sec$^{-1}$ (in the right panel).  Fig. \ref{fig3} shows
the case for the D2 distribution function for a value of $\rho_0$
equal to its maximal value, $\rho_0 = {\rho_0}^{\rm max}$; thus,
$\rho_0 = 0.50$ GeV cm$^{-3}$ for $v_0 = 170$ km sec$^{-1}$ (in the
left panel) and $\rho_0 = 1.27$ GeV cm$^{-3}$ for $v_0 = 270$ km
sec$^{-1}$ (in the right panel).  The values for ${\rho_0}^{\rm min}$
and ${\rho_0}^{\rm max}$ employed here are taken from Table III of
Ref. \cite{bcfs}.  A further example of annual-modulation regions,
corresponding to the {standard} DF, the cored--isothermal sphere A0
with $\rho_0 = 0.34$ GeV cm$^{-3}$ and $v_0 = 220$ km sec$^{-1}$, will
be given in Fig. \ref{th-fig3} in Sect. \ref{sec:comp}, where we
compare the LNM with the experimental results.  In all figures the
escape velocity has been maintained at the fixed value: 650 km/s.  Of
course, the present existing uncertainties affecting the knowledge of
the escape velocity -- as well as other uncertainties not included
here -- would significantly modify/extend the allowed regions.

The three (colored) hatched regions in Figs. \ref{fig1}, \ref{fig2} and \ref{fig3} denote the DAMA annual 
modulation regions, under the hypothesis that the effect is due to a WIMP with a coherent interaction with nuclei 
and in 3 different instances:
i) without including the channeling effect ((green) vertically-hatched region),
ii) by including the channeling effect according to Ref. \cite{chan} ((blue) horizontally-hatched region), and 
iii) without the channeling effect but using an energy--dependent Na and I quenching factors as established by the 
procedure given in Ref. \cite{tretyak} ((red) cross-hatched region).
It is worth noting that, depending on the possible amount of blocking effect in NaI(Tl) with respect to the modeling 
used in Ref. \cite{chan}, the channeled (blue) region will span the domain between the present channeled 
region and the unchanneled one. Moreover, the availability of quenching factor values not integrated over a large 
energy interval can also play a relevant role.

All these DAMA regions have been investigated here in some specific
cases by adopting a procedure that allows to put into evidence -- to
some extent -- the uncertainties on the quenching factors and on the
nuclear form factors: by considering the mean values of the parameters
of the used nuclear form factors and of the quenching factors of
Ref. \cite{Psd96} (case A); by varying the mean values of those
quenching factors up to +2 times the errors quoted there and the
nuclear radius, $r_n$, and the nuclear surface thickness parameter,
$s$, in the SI form factor from their central values down to $-20\%$
(case B); by fixing the Iodine nucleus parameters at the values of
case B, while for the Sodium nucleus one considers the quenching
factor at the lowest value measured in the literature and the nuclear
radius, $r_n$, and the nuclear surface thickness parameter, $s$, in
the SI form factor from their central values up to $+20\%$ (case C).

The DAMA regions have been obtained by superposition of the three
regions corresponding to the cases A, B, and C.  These regions
represent -- as in some previous DAMA publications -- the domain where
the likelihood-function values differ more than 7.5 $\sigma$ from the
null hypothesis (absence of modulation). This choice allows both a
direct superposition of the obtained results for both Na and I target
nuclei (which case by case can have different levels of the
corresponding minimum value of the likelihood function) and a very
high C.L. requirement.

In the same figures Figs.\ref{fig1},\ref{fig2},\ref{fig3} the allowed
regions by the CoGeNT experiment \cite{cogent} (under the same adopted
framework) are reported, assuming for simplicity for the Ge a fixed
value of 0.2 for the quenching factor and a Helm form factor with
fixed parameters.  In particular, the CoGeNT regions have been
obtained by fitting the measured modulation amplitudes with the WIMP
expectation ($S_m$) and using the 0.45--3.15 keV energy region ($R$ in
the following) of the energy spectrum as a constraint.  The $\chi^2$
function is:
\begin{equation}
\chi^2 = \sum_{k=1,2} \frac{(S_{m,k} - A_k)^2}{\sigma_{A,k}^2} +
\sum_R \frac{(S_{0,k} - r_k)^2}{\sigma_k^2} \Theta (S_{0,k} - r_k)\, ,
\label{eq:coge}
\end{equation}
where $A_k$ and $\sigma_{A,k}$ are the modulation amplitudes and their
errors in the two considered energy bins; $r_k$ and $\sigma_k$
are the rates and their errors in the $k$ energy bin. The $\Theta$
Heaviside function occurs in the second term to account for the
constraint of the rate in those energy bins ($R$).  In particular, we
derived from Ref. \cite{cogent} the following modulation amplitudes:
$A(0.5-0.9)$ keV $= (0.91 \pm 0.61)$ cpd/kg/keV; $A(0.5-3.0)$ keV $=
(0.45 \pm 0.18)$ cpd/kg/keV.  Thus, we consider in eq. \ref{eq:coge}
$A_{k=1} = A(0.5-0.9)$ keV and we infer $A_{k=2} = A(0.9-3.0)$ keV $=
(0.36 \pm 0.18)$ cpd/kg/keV.
The values of the modulation amplitudes have been obtained here
under the assumption that the period and the phase of the modulation
 are fixed at their nomimal values of 1 year and June $2^{\rm nd}$.
If one allows the phase and the period to be
free parameters, the ensuing modulation amplitudes occur to
be larger, but still compatible within the quoted errors.

The (non-hatched) regions denoted by (black) solid contours in
Figs. \ref{fig1}, \ref{fig2} and \ref{fig3} denote the allowed regions
by the CoGeNT experiment; such regions contain configurations whose
likelihood--function values differ more than 1.64 $\sigma$ from the
null hypothesis (absence of modulation). This corresponds roughly to
90\% CL far from zero signal.  In fact due to the presently more
modest C.L. (about 2.9 $\sigma$) of the CoGeNT result with respect to
the 9 $\sigma$ C.L. of the DAMA/NaI and DAMA/LIBRA evidence for DM
particles in the galactic halo, obviously no region is found if the
stringent 7.5 $\sigma$ from absence of modulation is required as for
the DAMA cases; thus, it will be very interesting to see future CoGeNT
data releases with increased significance.  Anyhow, all the examples
given here, as well as the proper inclusion of possible uncertainties
in the assumptions adopted for CoGeNT, and additional
accounting of other uncertainties, offer a substantial agreement
between the two experiments (as well as with some preliminary possible
positive hint by CRESST discussed at Conferences so far \cite{cresst},
which is not addressed here) towards a low mass candidate.

From Figs. \ref{fig1}, \ref{fig2} and \ref{fig3} we see that in all instances the DAMA and the CoGeNT 
regions agree quite well, over ranges of $\xi\sigma_{\rm scalar}^{(\rm nucleon)}$ and $m_{\chi}$ somewhat wider
as compared to those derived for instance in Refs. \cite{cogent,kelso}. The gross features in the comparative positions 
of the various regions in our Figs. \ref{fig1}, \ref{fig2} and \ref{fig3} are easily understood in terms 
of the specific values of the DF parameters employed. Further statistics in the CoGeNT experiment will be useful in 
pinning down more precisely the common domains for the two annual--modulation experiments.

\section{The Light Neutralino Model}
\label{sec:lnm}

Now we discuss how the results reported in the previous Section are well fitted by the light neutralinos which arise
within the model introduced in Ref. \cite{lowneu} and developed in the papers of Ref.  \cite{sum}. Lately,
 this model, denoted as  Light Neutralino Model (LNM), was updated in Refs. \cite{discussing,impact}
 to take into account  recent constraints on supersymmetric models derived at accelerators and B--factories.

\subsection{Main features of the LNM}
\label{sec:main}

The LNM is an effective MSSM scheme at the electroweak scale with the
following independent parameters: $M_1, M_2, M_3, \mu, \tan\beta, m_A,
m_{\tilde q}, m_{\tilde l}$ and $A$.  Notations are as follows:
$M_1$, $M_2$ and $M_3$ are the U(1), SU(2) and SU(3) gaugino masses
(these parameters are taken here to be positive), $\mu$ is the Higgs
mixing mass parameter, $\tan\beta$ the ratio of the two Higgs
v.e.v.'s, $m_A$ the mass of the CP--odd neutral Higgs boson, 
  $m_{\tilde{q}}$ is a squark soft--mass common to the all families,
$m_{\tilde l}$ is a slepton soft--mass common to all sleptons, and $A$
is a common dimensionless trilinear parameter for the third family,
$A_{\tilde b} = A_{\tilde t} \equiv A m_{\tilde q}$ and $A_{\tilde
  \tau} \equiv A m_{\tilde l}$ (the trilinear parameters for the other
families being set equal to zero). We recall that in
  Ref. \cite{impact} the possibility of a splitting between the squark
  soft--mass common to the first two families and that of the third
  family was considered. This allows to reduce the fine--tuning in the
  parameters that can be induced by the interplay between the
  constraint from the $b\rightarrow s \gamma$ decay and those from SUSY
  searches at the LHC.

The linear superposition of bino $\tilde B$, wino $\tilde W^{(3)}$
and of the two Higgsino states $\tilde H_1^{\circ}$, $\tilde
H_2^{\circ}$ which defines the neutralino state of lowest mass $m_{\chi}$
will be written here as:

\begin{equation}
\chi \equiv a_1 \tilde B + a_2 \tilde W^{(3)} +
a_3 \tilde H_1^{\circ} + a_4  \tilde H_2^{\circ}.
\label{neutralino}
\end{equation}

Since no
gaugino-mass unification at a Grand Unified scale is assumed in our LNM
(at variance with one of the major assumptions in mSUGRA), in this model
the neutralino mass is not bounded by the lower limit
$m_{\chi} \gsim$ 50 GeV that is commonly derived in mSUGRA schemes
from the LEP lower bound on the chargino mass (of about 100 GeV).
In Refs. \cite{lowneu,lowneudama,sum,discussing} it is shown that, if
R--parity is conserved, a light neutralino
({\it i.~e.} a neutralino with $m_{\chi} \lsim$ 50 GeV) is a very
interesting candidate for cold dark matter (CDM), due to its
relic abundance and its relevance in the interpretation of current
experiments of search for relic particles; it is also
shown there that
a lower bound $m_{\chi} \gsim$ 7--8 GeV is obtained from the
cosmological upper limit on CDM. The compatibility of these results
with all experimental searches for direct or indirect evidence of
SUSY (prior to the first physics results of LHC) and with other
precision data that set constraints on possible effects due to
supersymmetry
is discussed in detail in Ref. \cite{discussing}.
The viability of very light neutralinos in terms of various  constraints
from collider data, precision observables and rare meson decays is also considered 
in Ref. \cite{Dreiner:2009ic}. Perspectives for investigation of these neutralinos at LHC  are analyzed in Ref. \cite{lhc1} and prospects for a very accurate mass measurement at ILC in Ref. \cite{Conley:2010jk}.

In the present section we essentially recall the main properties of the light
neutralinos within the LNM, as derived in Refs. \cite{lowneu,lowneudama,sum,discussing},
and  which are relevant for the discussion of the experimental
results of Refs. \cite{dama2010,cogent}.

In the regime of light neutralinos the lower limit on
the mass $m_{\chi}$, obtained from the requirement that
its relic abundance does not exceed  the observed upper bound for cold dark matter (CDM), {\it i.e.}
$\Omega_{\chi} h^2 \leq (\Omega_{CDM} h^2)_{\rm max}$, can be expressed
analytically in terms of the relevant SUSY parameters.
In this concern, it is convenient to distinguish between   two scenarios.
 The first one is denoted as {Scenario $\mathcal{A}$},
 and its main features are: i) $m_A$ is light,
  90 GeV $\leq m_A \lsim (200-300)$ GeV
 (90 GeV being the lower bound from LEP searches);
  ii)  $\tan \beta$ is large: $\tan \beta$ = 20--45,
  iii) the  $\tilde {B} - \tilde H_1^{\circ}$ mixing needs to be sizeable, which in
  turn implies small values of $\mu$: $|\mu| \sim (100-200)$ GeV.
  In this scenario  the dominant contribution to
  the annihilation cross--section of a pair
of neutralinos, $\sigma_{\rm ann}$ (which establishes the size of the
neutralino relic abundance) is provided by the $A$--exchange in
the $s$ channel of the
annihilation process $\chi + \chi \rightarrow \bar{b}+b$, thus
  the lower bound on $m_{\chi}$ is given by:

\begin{widetext}
\begin{equation}
m_{\chi}~ \frac{[1-m_b^2/m_\chi^2]^{1/4}}{[1-(2m_{\chi})^2/m_A^2]}
\gsim 7.4 ~ {\rm GeV} \left(\frac{m_A}{90 \; {\rm GeV}} \right)^2 \left(\frac{35}{\tan \beta}\right)
\left(\frac{0.12}{a_1^2 a_3^2}\right)^{\frac{1}{2}}
\left(\frac{0.12}{(\Omega_{CDM} h^2)_{\rm max}}\right)^{\frac{1}{2}},
\label{ma}
\end{equation}
\end{widetext}

\noindent
where $m_b$ is the mass of the $b$ quark.

When $m_A \gsim (200-300)$ GeV, the cosmological upper bound
on the neutralino relic abundance  can be satisfied by a pair annihilation process
which proceeds through an efficient stau--exchange contribution (in the
{\it t, u} channels). This requires that: (i) the stau mass
$m_{\tilde{\tau}}$ is sufficiently light, $m_{\tilde{\tau}} \sim$ 90
GeV (notice that the current experimental limit is $m_{\tilde{\tau}} \sim$ 87 GeV) and (ii) $\chi$ is a very pure Bino ({\it i.e.} $(1 -
a^2_1) \sim {\cal O}(10^{-2})$). Thus, one is lead to a
Scenario {$\mathcal{B}$, identified by the following sector of the supersymmetric
parameter space: $M_1 \sim$ 25 GeV, $|\mu| \gsim$ 500 GeV, $\tan \beta
\lsim$ 10; $m_{\tilde{l}} \gsim (100 - 200)$ GeV, $-2.5 \lsim A \lsim +2.5$.
As derived in Ref.  \cite{lowneu,lowneudama,sum,discussing}, in this scenario the cosmological bound
$\Omega_{\chi} h^2 \leq (\Omega_{CDM} h^2)_{\rm max}$ provides
the lower bound  $m_{\chi} \gsim 22$ GeV  \cite{hooper_bel},
whose scaling law in terms of  the stau mass
and $(\Omega_{CDM} h^2)_{\rm max}$ is approximately given by:

\begin{widetext}
\begin{equation}
m_{\chi} [1-m_{\tau}^2/m_\chi^2]^{1/4} \gsim
22 \; {\rm GeV} \;
\left( \frac{m_{\tilde{\tau}}}{90 \; {\rm GeV}} \right)^2 \left(\frac{0.12}{(\Omega_{CDM} h^2)_{\rm max}}\right).
\label{tau}
\end{equation}
\end{widetext}

In  {Scenario $\mathcal{A}$}  the neutralino--nucleon cross--section
$\sigma_{\rm scalar}^{(\rm nucleon)}$
is dominated by  the interaction process  due to the exchange of the lighter CP--even
neutral Higgs boson $h$, whose mass $m_h$ has a numerical value very close to
$m_A$;  then  $\sigma_{\rm scalar}^{(\rm nucleon)}$ is expressible as:

\begin{widetext}
\begin{equation}
\sigma_{\rm scalar}^{(\rm nucleon)} \simeq 5.3  \times 10^{-41} \; {\rm cm^2} \;
\left(\frac{a_1^2 a_3^2}{0.13} \right)
\left(\frac{\tan \beta}{35} \right)^2
\left(\frac{90 \; {\rm GeV}}{m_h} \right)^4
\left(\frac{g_d}{290 ~ {\rm MeV}} \right)^2,
\label{sel2}
\end{equation}
\end{widetext}
where $g_d$ is the dominant coupling in the interaction of the Higgs boson
 $h$ with the $d$--type quarks,

\begin{equation}
g_d \equiv [m_d \langle N|\bar{d} d |N\rangle + m_s \langle N|\bar{s} s |N\rangle +
m_b \langle N|\bar{b} b |N\rangle],
\label{eq:gd}
\end{equation}
and $\langle N|\bar{d} d |N\rangle$ denotes the scalar density of a
generic quark $q$ inside the nucleon. In this expression we have used
as {\it reference} value for $g_d$ the value $g_{d,\rm ref} = 290$ MeV
employed in our previous papers \cite{sum}. We recall that this
quantity is affected by large uncertainties \cite{uncert2} with
$\left({g_{d,\rm max}}/{g_{d,\rm ref}}\right)^2 = 3.0$ and
$\left({g_{d,\rm min}}/{g_{d,\rm ref}}\right)^2 = 0.12$, a fact that
directly transforms in the same amount of uncertainty on the coherent
scattering cross section.

Since, as mentioned in Sect. \ref{sec:region},  we wish to consider also situations where relic
neutralinos only provide a fraction of the CDM abundance,
the relevant quantity we will compare with the experimental results
is not simply $\sigma_{\rm scalar}^{(\rm nucleon)}$, but rather $\xi
\sigma_{\rm scalar}^{(\rm nucleon)}$. The factor
$\xi = \rho_{\chi} / \rho_0$ is calculated here according to the {\it rescaling}
recipe
 $\xi = {\rm min}\{1, \Omega_{\chi} h^2/(\Omega_{CDM} h^2)_{\rm min}\}$  \cite{gaisser}, where
$(\Omega_{CDM} h^2)_{\rm min}$ is the minimal value to be assigned to the
relic abundance of CDM.

It is remarkable that for
 neutralino configurations, {\it whose relic abundance stays in the
cosmological range for CDM} ({\it i.e.} $(\Omega_{CDM} h^2)_{\rm min} \leq \Omega_{\chi} h^2 \leq (\Omega_{CDM} h^2)_{\rm max}$ with
$(\Omega_{CDM} h^2)_{\rm min} = 0.098$ and  $(\Omega_{CDM} h^2)_{\rm max} = 0.12$) and pass all particle--physics constraints,  the elastic neutralino--nucleon cross--section can be cast as  \cite{lowneu,lowneudama,sum,discussing}:

\begin{widetext}
\begin{equation}
\sigma_{\rm scalar}^{(\rm nucleon)} \simeq (2.7 - 3.4) \times 10^{-41} \; {\rm cm^2}  \;
\left(\frac{g_d}{290 ~ {\rm MeV}} \right)^2
\frac{[1-(2m_{\chi})^2/m_A^2]^2}{(m_{\chi}/(10 \; {\rm GeV})^2 \; [1- m_b^2/m_\chi^2]^{1/2}}.
\label{bound}
\end{equation}
\end{widetext}
\noindent

Notice that this formula provides an evaluation of  $\sigma_{\rm scalar}^{(\rm nucleon)}$
simply in terms of the neutralino mass $m_{\chi}$.
Specific SUSY parameters such as $\tan \beta$ or $\mu$ do not appear explicitly, since
these parameters have been reabsorbed by the introduction of the relic abundance
(this is because here the annihilation amplitude is related to the elastic-scattering
amplitude by crossing symmetry). The remaining dependence on the mass of the interaction mediator
$m_A \simeq m_h$ is only marginal, due to the small values of $m_{\chi}$ considered here.

 Eq.(\ref{bound}) is of particular interest in establishing the range of values for
 $\sigma_{\rm scalar}^{(\rm nucleon)}$ in terms of the neutralino mass. The numerical range
 in front of Eq.(\ref{bound}) follows from the requirement that relic neutralinos have
 an abundance in the cosmological range for CDM.
 The crucial factor of uncertainties in $\sigma_{\rm scalar}^{(\rm nucleon)}$
 is related to QCD properties through
 the coupling $g_d$.  It is however worth recalling that the range of the neutralino
 mass depends on the lower bound on  $m_{\chi}$ which is explicitly given in terms of the SUSY parameters
 in Eq.(\ref{ma}).
These properties will also show up later in the figures displaying the scatter plots
for $\sigma_{\rm scalar}^{(\rm nucleon)}$.

\subsection{Constraints on SUSY parameters from early searches at the LHC}
\label{sec:lhc}

\begin{figure}[t]
\includegraphics[width=1.0\columnwidth]{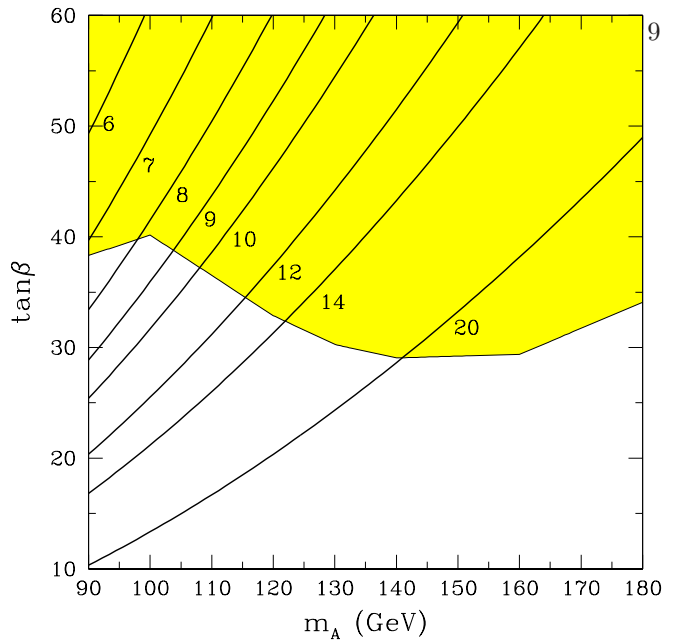}
  \caption{ Upper bounds in the $m_A$ -- $\tan\beta$ plane, derived in
    Ref. \cite{baglio} from searches of the neutral Higgs boson
    decaying into a tau pair at LHC \cite{cmshiggs}. The disallowed
    domain is the (yellow) shaded region.  The solid bold lines
    labeled by numbers denote the cosmological bound $\Omega_\chi h^2
    \leq (\Omega_{CDM} h^2)_{\rm max}$ for a neutralino whose mass is
    given by the corresponding number (in units of GeV), as obtained
    by Eq. (\ref{ma}) with $(\Omega_{CDM} h^2)_{\rm max}=0.12$. For
    any given neutralino mass, the allowed region is above the
    corresponding line.  }
\label{th-fig1}
\end{figure}

\begin{figure*}[t]
\includegraphics[width=0.45\textwidth] {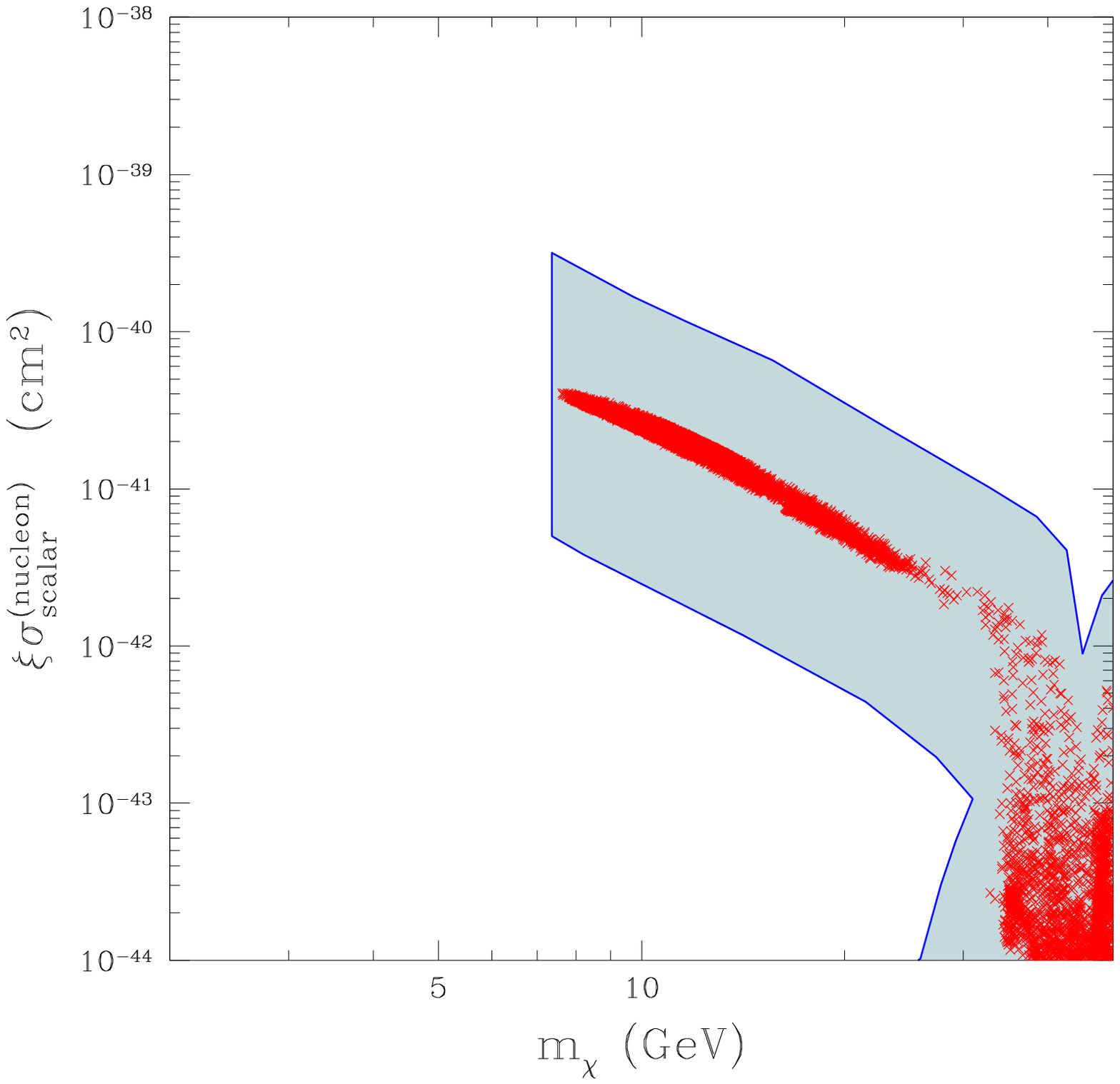}
\includegraphics[width=0.45\textwidth] {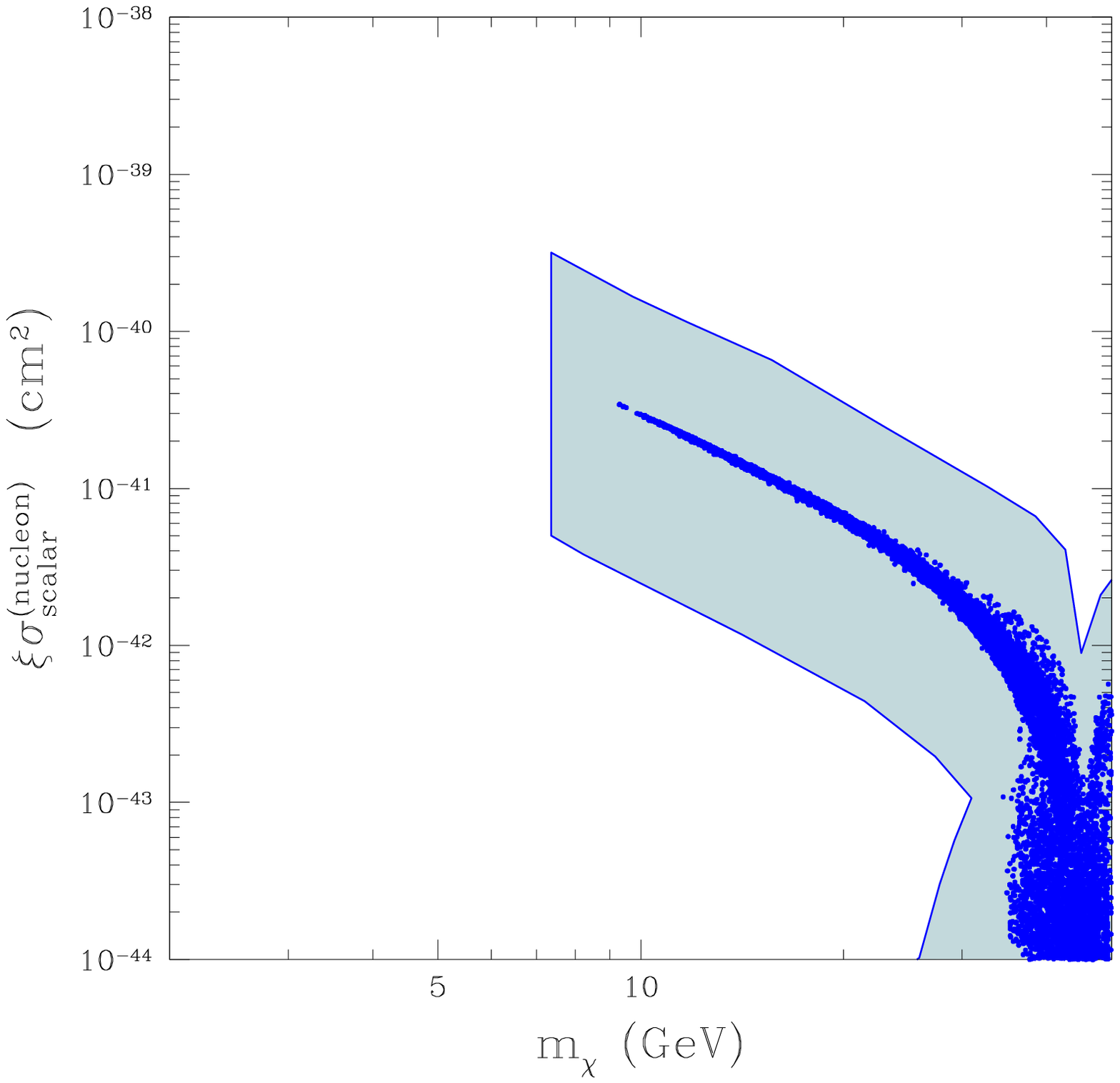}
\caption{
Scatter plot for $\xi \sigma_{\rm scalar}^{(\rm nucleon)}$ as a function of
the neutralino mass for $g_{d,\rm ref}$ = 290 MeV.
The left panel displays by  (red) crosses SUSY configurations with a neutralino
relic abundance which matches the WMAP cold dark
matter amount (0.098 $\leq \Omega_{\chi} h^2 \leq$ 0.122) whereas
the right panel displays by (blue) dots
the  configurations where the neutralino is subdominant
($ \Omega_{\chi} h^2 <$ 0.098).
The (light-blue) flag--like region denotes the
extension of the scatter plot upwards and
downwards, when the hadronic uncertainties are included.}
\label{th-fig2}
\end{figure*}

In Ref. \cite{impact} the possible impact of some early analyses by
the CMS and ATLAS Collaborations at the LHC on the LNM was investigated.
The data considered there consisted in the results of searches for supersymmetry
 in proton--proton collisions at a center--of--mass energy of 7 TeV
with an integrated luminosity of 35 ${\rm pb}^{-1}$  \cite{cms}, {\it i.e.}
the results of the CMS Collaboration for events with jets and missing transverse energy  \cite{cms}, and those of  the ATLAS Collaboration
 by studying
 final states containing jets, missing transverse energy, either with a isolated lepton (electron or muon)  \cite{atlas}
 or without final leptons  \cite{atlas2}.
 Both signatures would be significant
of processes due to the production in pairs of
squarks and gluinos, subsequently decaying into quarks, gluons, other standard-model (SM)
particles and a neutralino (interpreted as the lightest supersymmetric particle (LSP)) in
a R--parity conserving SUSY theory. As reported in Refs. \cite{cms,atlas}  the data appeared
to be consistent with the expected SM backgrounds; thus lower bounds were derived
on the squark and gluino masses which are sizeably higher than
the previous limits established by the experiments D0  \cite{D0} and CDF  \cite {CDF} at the Tevatron.

  These data were employed in Ref. \cite{impact} to determine the
  relevant lower bounds on the squark masses and the gluino mass $M_3$
  within the LNM and their ensuing possible impact on the value of the
  lower bound on the neutralino mass. It was proved there that the
  data of Refs. \cite{cms,atlas} do not imply a modification of the
  lower bound $m_{\chi} \gsim$ 7-8 GeV for the LNM, when the common
  squark mass for the first two families $m_{\tilde{q}_{12}}$ and the
  one for the third family $m_{\tilde t}$ are independent parameters
  with $m_{\tilde{q}_{12}} > m_{\tilde t}$, or, in case of a full
  degeneracy of the squark masses over the 3 families (as considered
  in the present paper), when $M_3 \gsim$ (1.5 - 2) TeV. Otherwise, in
  the case of a full squark--mass degeneracy ($m_{\tilde{q}_{12}} =
  m_{\tilde t}\equiv m_{\tilde{q}}$) the lower bound on $m_{\chi}$
  varies as a function of the gluino mass $M_3$, from the value of
  7--8 GeV for $M_3 \gsim 2$ TeV to about 12 GeV for $M_3 \simeq$ 600
  GeV (see Fig. 5 of Ref.  \cite{impact} for details). In particular,
  the gluino mass enters in the calculation of observables for the
  relic neutralino only at the loop level (through radiative
  corrections of Higgs couplings), so within the LNM $M_3$ is very
  weakly correlated to the other parameters. In order to reduce the
  number of parameters, in the present analysis we choose to decouple the
  gluino mass assuming $M_3=$2 TeV. In this case LHC data imply the
  lower bound $m_{\tilde{q}}\gsim$ 450 within the LNM
  \cite{impact}. In the following we will impose this constraint in
  our numerical analysis.

Now, we proceed to a discussion of the new results presented by the CMS Collaboration on a search for
neutral SUSY Higgs bosons decaying in tau pairs at a center--of--mass energy of 7 TeV with an integrated
luminosity of 36 ${\rm pb}^{-1}$  \cite{cmshiggs}. Since no excess is observed in the tau-pair invariant--mass spectrum, upper limits on the Higgs--boson production cross section times the
branching ratio to tau pairs are placed. These limits are then converted into upper bounds for the SUSY
parameter $\tan \beta$ as a function of the pseudoscalar Higgs--boson mass $m_A$ in a particular MSSM benchmark.
The ensuing  disallowed region in the plane $m_A - \tan \beta$ turns out to be  considerably
larger than the one previously derived at the Tevatron (see for instance Ref. \cite{tevnp}).

However, in Ref. \cite{baglio} it has been shown that, when all the theoretical uncertainties involved in
the derivation of the previous bounds
on the Higgs--boson production cross section times the
branching ratio to tau pairs
are appropriately taken into account, the limits on the SUSY parameters
reported in Refs. \cite{cmshiggs,tevnp}  are significantly relaxed.

We display in Fig. \ref{th-fig1} the region disallowed in the plane
($\tan \beta - m_A$) from the results of Refs. \cite{cmshiggs}, as
derived in the analysis of Ref. \cite{baglio}.  In this figure we also
show the lines corresponding to fixed values of the neutralino mass in
the LNM. Thus we see that the CMS upper bounds of Ref. \cite{cmshiggs}
do not modify the value of the neutralino--mass lower bound $m_{\chi}
\gsim$ 7--8 GeV, previously derived in Refs.
\cite{sum,discussing}. This result also follows directly from the
analytic expression of Eq. (\ref{ma}) for the lower limit on $m_\chi$.

The predictions of the LNM for the cross-section $\sigma_{\rm
  scalar}^{(\rm nucleon)}$ were already anticipated in the analytic
expressions of Eqs.(\ref{sel2})--(\ref{bound}). Now we give the
numerical values for the quantity $\xi \sigma_{\rm scalar}^{(\rm
  nucleon)}$ which will finally be confronted with the experimental
results.  Fig. \ref{th-fig2} provides the scatter plots of this
quantity for the neutralino configurations which pass all the
constraints previously discussed in this Section.
In particular, in our scan of the
  LNM the following ranges of the parameters are adopted: $10 \leq
  \tan \beta \leq 50$, $105 \, {\rm GeV} \leq \mu \leq 1000 \, {\rm
    GeV}$, $5 \, {\rm GeV} \leq M_1 \leq 50 \, {\rm GeV}$, $100 \,
  {\rm GeV} \leq M_2 \leq 2500 \, {\rm GeV}$, $450 \, {\rm GeV} \leq
  m_{\tilde q} \leq 3000 \, {\rm GeV }$, $115 \, {\rm GeV} \leq
  m_{\tilde l} \leq 3000 \, {\rm GeV }$, $90\, {\rm GeV }\leq m_A \leq
  1000 \, {\rm GeV }$, $-3 \leq A \leq 3$.

The left panel refers to SUSY configurations with a neutralino relic
abundance which matches the WMAP cold dark matter amount (0.098 $\leq
\Omega_{\chi} h^2 \leq$ 0.122) whereas the right panel displays by
(blue) dots the configurations where the neutralino is subdominant ($
\Omega_{\chi} h^2 <$ 0.098). In both panels, the flag--like region
denotes the extension of the scatter plots upwards and downwards, when
the hadronic uncertainties are included.

\begin{figure}[t]
\includegraphics[width=1.0\columnwidth]{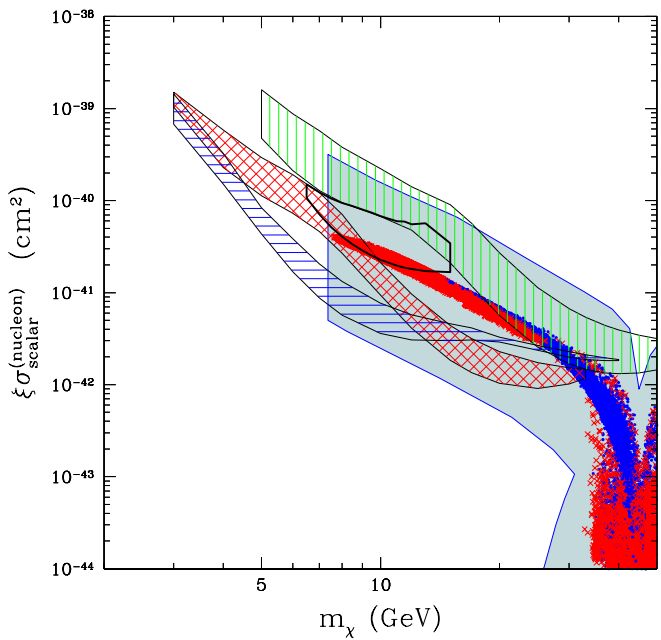}
  \caption{$\xi \sigma_{\rm scalar}^{(\rm nucleon)}$ as a function of
    the neutralino mass.  The experimental annual-modulation regions
    are obtained as explained in the caption of Fig.1, except that
    here the used DF is an isothermal sphere with the following values
    for the parameters: $\rho_0 = 0.34$ GeV cm$^{-3}$, $v_0 = 220$ km
    sec$^{-1}$, $v_{esc} = 650$ km sec$^{-1}$.  The theoretical
    scatter plot displays the whole sample of neutralino
    configurations: (red) crosses denote SUSY configurations with a
    neutralino relic abundance which matches the WMAP cold dark matter
    amount (0.098 $\leq \Omega_{\chi} h^2 \leq$ 0.122) while
    (blue) dots denote the configurations where the neutralino is
    subdominant ($ \Omega_{\chi} h^2 <$ 0.098) (these two sets of
    configurations were shown separately in Fig.5).  The scatter plot
    has been evaluated for $g_{d,\rm ref}$ = 290 MeV.  The
    (light-blue) flag--like region denotes the extension of the
    scatter plot upwards and downwards, when the hadronic
    uncertainties are included (see text).}
\label{th-fig3}
\end{figure}

\begin{figure}[t]
\includegraphics[width=0.45\textwidth] {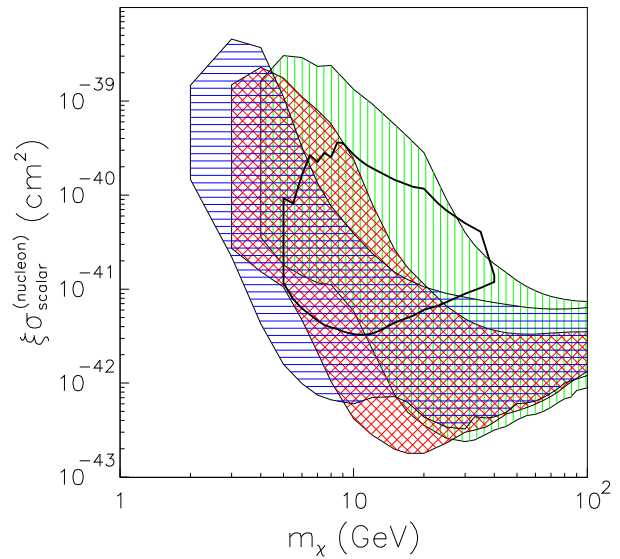}
  \caption{Regions in the $\xi\sigma_{\rm scalar}^{(\rm nucleon)}$ vs
    $m_{\chi}$ plane allowed by DAMA experiments in the three
    considered instances for the Na and I quenching 
        factors, including all the DFs considered in Ref. \cite{bcfs} and the same
  uncertainties as in Refs. \cite{RNC,ijmd} for a
    WIMP with a pure SI coupling.  The hatchings (and colours) of the allowed regions
    are the same as those in Fig. \ref{fig1}.  These regions represent
    the domain where the likelihood--function values differ more than
    7.5 $\sigma$ from the null hypothesis (absence of modulation).  It
    is worth noting that, depending on other possible uncertainties
    not included here, the channeled (blue) horizontally--hatched region could span the
    domain between the present channeled region and the unchanneled
    one.  The allowed region obtained for the CoGeNT
    experiment, including the same astrophysical models as in
    Ref. \cite{RNC,ijmd} and assuming for simplicity a fixed value for the Ge quenching
    factor and a Helm form factor with fixed
    parameters, is also reported and denoted by a (black) thick solid
    line.  This region is meant to include configurations whose
    likelihood--function values differ more than 1.64 $\sigma$ from
    the null hypothesis (absence of modulation). This corresponds
    roughly to 90\% CL far from zero signal. See text.  }
\label{fig5}
\end{figure}

\section{Comparison with experimental results}
\label{sec:comp}

The predictions for $\xi \sigma_{\rm scalar}^{(\rm nucleon)}$ for light neutralinos within the LNM, as depicted in
Fig. \ref{th-fig2}, fall clearly in the region of interest of the present annual--modulation results as reproduced in Figs. \ref{fig1},\ref{fig2},\ref{fig3} and in Fig. \ref{fig5} (to follow).
For a more specific comparison among experiments and theory we employ here, as a reference DF, the {\it standard} isothermal sphere with parameters: $\rho_0 = 0.34$ GeV cm$^{-3}$, $v_0 = 220$ km sec$^{-1}$,
$v_{esc} = 650$ km sec$^{-1}$.   This choice is not meant to attribute to this particular DF a privileged role over other DFs, but is done simply for convenience, to conform to the most commonly employed form of the DF. The experiment--theory comparison is therefore displayed in Fig. \ref{th-fig3}. The features of this figure confirm that the conclusions drawn in
Ref. \cite{discussing} are even reinforced, when, as done in the present paper, the new annual-modulation results by CoGeNT are included; specifically: i) the light neutralino population agrees with the DAMA/LIBRA annual modulation data over a wide range of light neutralinos: 7--8 GeV $\lsim m_{\chi} \lsim$ 50 GeV, ii) this population is also in agreement with the data of
CoGeNT in a range of the neutralino mass somewhat restricted to the lower masses: 7--8 GeV $\lsim m_{\chi} \lsim$ (15-20) GeV.

It is worth recalling that also the data of CDMS \cite{cdms}, and CRESST \cite{cresst}, should their reported excesses be
significant of real DM signals, would fall in a domain of
the $\xi \sigma_{\rm scalar}^{(\rm nucleon)}$-$m_{\chi}$ plane
 overlapping with the DAMA and CoGeNT regions.

\section{Conclusions}
\label{sec:con}

The long--standing model--independent annual modulation signal measured by the DAMA Collaboration
for a total exposure of 1.17 ton $\times$ year and a confidence level of 8.9 $\sigma$
with a NaI(Tl) detector  \cite{dama2010}
has been comparatively examined with the new results by the CoGeNT experiment
 \cite{dama2010}
which shows a similar behavior with a statistical significance of about 2.86 $\sigma$.
The annual modulation measured in these two  experiments is an effect expected
because of the relative motion of the Earth
with respect to the relic particles responsible for the dark matter in the
galactic halo  \cite{freese}. The underlying physical process can be due to a variety
of interaction mechanisms of relic particles with the detector materials  \cite{dama2008}.
Here we have limited our analysis  to the case where the signal would be caused
by nuclear recoils induced by elastic coherent interactions of the target nuclei
with the DM particles.

The ensuing physical regions in the plane of the DM--particle mass
versus the DM--particle -- nucleon cross--section have been  derived for a variety of
DM distribution functions in the galactic halo and by taking into account the
impact of various experimental uncertainties.

The phase--space distribution of DM particles in the halo is still subject of
extensive astrophysical investigations, with the possible presence of
unvirialized components (see, for instance, Ref. \cite{spergel}). Here we have
selected a few samples of DFs, selected among those discussed in
Ref. \cite{bcfs}, from the isothermal sphere to the Jaffe DF
 \cite{jaffe} to a triaxial one  \cite{evans}.

 We have examined in details to what extent the major experimental
 uncertainties, most notably those related to the quenching factors
 and the channeling effect, affect the derivation of the
 annual--modulation physical regions. It is shown that the DAMA and
 the CoGeNT regions agree well between each other independently of the
 specific analytic form of the DFs considered here, considering also
 that some existing uncertainties have not been taken into account for
 the CoGeNT allowed regions. For completeness, Fig. \ref{fig5} shows
 the DAMA allowed regions in the three considered instances for the Na
 and I quenching 
 factors when including all
  the DFs considered in Ref. \cite{bcfs} and the same  uncertainties as in
  Ref. \cite{RNC,ijmd}. The allowed region obtained for the
 CoGeNT experiment, including the same astrophysical models as in
 Ref. \cite{RNC,ijmd} and assuming for simplicity a fixed value for the Ge quenching
 factor and a Helm form factor with fixed parameters,
 is also reported in Fig. \ref{fig5} (solid line); it fully overlaps
 the DAMA allowed regions. The inclusion of other uncertainties on
 parameters and models (see for example Refs. \cite{RNC,ijmd}) would
 further enlarge these regions.

In this paper, we have finally discussed a specific particle--physics
realization, the Light Neutralino Model, where neutralinos
with masses in the tens of GeV range naturally arise.
This supersymmetric model,
which was already shown  \cite{discussing} to be successful in fitting the DAMA annual modulation results
 \cite{dama2010} as well
as the (unmodulated) CoGeNT  \cite{cogent2010}, the CDMS  \cite{cdms} and the CRESST  \cite{cresst} excesses, is shown here to
agree quite well also with the most recent CoGeNT annual-modulation data  \cite{cogent}. Notice that the
LNM discussed here satisfies all available experimental particle--physics constraints,
including the most recent results from CMS and ATLAS at the CERN Large Hadron Collider.
Confirmation of the validity of the SUSY model discussed in the present paper rests on the
possibility of a positive evidence of light neutralinos in further running of LHC  \cite{lhc1}.

\medskip
{\bf Note Added.}
In this Note we comment on two preprints that appeared after the 
     submission of the present paper.

     The viability of an MSSM to obtain a neutralino-nucleon elastic 
     cross section with a size relevant for the DAMA and CoGeNT annual
     modulation data, for light neutralino masses, is questioned in the 
     preprint of Ref. \cite{cumber}. We note that in the MSSM scheme 
     employed in Ref. \cite{cumber} the squark masses are all set at 1
     TeV. From the properties discussed in detail in Refs.
     \cite{discussing,impact} it is clear that, taking all the squark 
     masses at this value generates tension between the $b \rightarrow s  
     + \gamma$ and the $B_s \rightarrow \mu^+ + \mu^-$ constraints,  and 
     thus precludes low values of the Higgs-boson masses ({\it i.e} close 
     the their LEP lower bounds). This, in turn, disallows neutralino 
     masses $\lsim$ 15 GeV (see Fig. 2 of Ref. \cite{impact}). At 
     variance with the conclusions of Ref. \cite{cumber}, in the present 
     paper it is shown that an appropriate MSSM scheme fits the DAMA and 
     CoGeNT annual modulation results quite well in force of the 
     properties spelled out in  Sect. \ref{sec:lnm}.

     Ref. \cite{fox} refers to our approach in the analysis of the CoGeNT 
     data as being ``somewhat unphysical'', since , according to Ref. 
     \cite{fox}, we would accept negative backgrounds. This is manifestly
     not the case, as can be easily understood by means of Eq.  
     (\ref{eq:coge}), which defines the statistical estimator we use in 
     our analysis. We explicitely enforce the bound arising from
     the total rate, in order not to accept modulation amplitudes which 
     would be incompatible with the measured total rate. The last term in 
     Eq. (\ref{eq:coge}) does, in fact, penalize the $\chi^2$ when the 
     calculated rate becomes exceedingly large, stastitically
     incompatible with the measured total rate. We therefore do not 
     accept negative backgrounds, contrary to the claim in Ref. 
     \cite{fox}.

\begin{acknowledgments}

The authors thank the DAMA Collaboration for making
available the data for this analysis.

A.B. and N.F. acknowledge Research Grants funded jointly by Ministero
dell'Istruzione, dell'Universit\`a e della Ricerca (MIUR), by
Universit\`a di Torino and by Istituto Nazionale di Fisica Nucleare
within the {\sl Astroparticle Physics Project} (MIUR contract number: PRIN 2008NR3EBK;
INFN grant code: FA51). S.S. acknowledges
support by NRF with CQUEST grant 2005-0049049 and by the Sogang
Research Grant 2010. N.F. acknowledges support of the spanish MICINN
Consolider Ingenio 2010 Programme under grant MULTIDARK CSD2009- 00064.

\end{acknowledgments}

\end{document}